\documentstyle[psfig]{l-aa}
\def\sane#1{\ifvmode{\leavevmode\hbox{#1}}\else{\hbox{#1}}\fi}
\def\sco{\sane{Sco X-1}}
\def\cyg{\sane{Cyg X-2}}
\def\gx5{\sane{GX 5$-$1}}
\def\agx17{\sane{GX 17+2}}
\def\bgx340{\sane{GX 340+0}}
\def\cgx349{\sane{GX 349+2}}
\def\about{$\sim$}

\newcommand{\mdot}{$\dot{M}$}
\newcommand{\Sz}{\mbox{S$_{\rm Z}$}}

\newcommand{\etal}{{\rm et~al.}}
\newcommand{\degree}{$^{\circ}$}
\newcommand{\rf}{\par\noindent\hangindent 15pt{}}

\begin{document}

\setcounter{page}{0}
\setcounter{figure}{0}
\psfull

\thesaurus{08
(03.13.2; 08.02.1; 08.09.2 Sco X$-$1; 08.09.2 Cyg X$-$2;
08.14.1; 13.25.5)}

\title{No QPO time lags from \sco as seen with EXOSAT: a comparison with \cyg.}

\author{
Stefan W.\, Dieters\inst{1,2}\thanks{{\em Present address}: 
SD50 NASA Marshall Space Flight Center, Huntsville, Alabama, AL 35812, U.S.A.,
e-mail: stefan.dieters@msfc.nasa.gov}
\and Brian A.\, Vaughan\inst{1,3} 
\and Erik Kuulkers\inst{1}\thanks{{\em Present Address}:
Space Research Organization Netherlands, Sorbonnelaan 2, 3584 CA Utrecht,
\& Astronomical Institute, Utrecht University, P.O Box 80000, 3507 TA Utrecht,
The Netherlands
} 
\and Frederick K. Lamb\inst{4}
\and Michiel van der Klis\inst{1}
}

\offprints{Stefan Dieters}

\institute{Astronomical Institute ``Anton Pannekoek'', University of
Amsterdam and Center for High Energy Astrophysics,
Kruislaan 403, 1098 SJ Amsterdam, The Netherlands.
\and 
 Max-Planck-Institut f\"ur Astrophsik, Postfach 15 23, D-85740
Garching b M\"{u}chen, Germany.
\and 
 Div. of Physics, Mathematics, and Astronomy, Caltech,
Pasadena, CA 91125, U.S.A.
\and
 Departments of Physics and Astronomy, University of Illinois at
Urbana-Champaign, 1110\,W.\,Green Street, Urbana, IL\,61801. U.S.A}

\maketitle
\date{Received 30 November 1997; accepted whenever}

\begin{abstract}

We have measured the phase-delay and rms amplitude spectra of \cyg\ and \sco. 
Using EXOSAT data from the normal branch of \cyg\ we confirm earlier (Ginga)
results, showing that at an energy near 6\,keV there is both a minimum in the
QPO rms amplitude spectrum and a 150\degree\ phase jump in the quasi-periodic
oscillations (QPO) phase-delay spectrum. 

Surprisingly, using EXOSAT and Ginga data, we find no evidence for a  phase
jump of this kind in the phase-delay spectrum of \sco\ on either the normal or
flaring branch. Upper limits (90\% confidence) of 42\degree\  can be set on
any phase jump in the energy range 2--10\,keV on the normal branch, and
88\degree\ on the flaring branch. The QPO rms amplitude spectrum of \sco\
increases steeply with energy on both the normal and flaring branches.  These
results suggest that the X-ray spectrum pivots about an energy of 
$\stackrel{<}{_{\sim}}$2\,keV or $\stackrel{>}{_{\sim}}$10\,keV or that normal
branch QPO of \sco\ does not have a pivoting spectrum. We discuss the
implications of these results in terms of the radiation-hydrodynamic model for
normal branch QPO.

\end{abstract}

\section{Introduction.}

After the discovery of quasi-periodic oscillations (QPO) in power density
spectra of the X-ray variability of \gx5\ (van der Klis \etal\  1985), QPO
were also discovered in  \sco\ (Middleditch \&\ Priedhorsky 1986) and \cyg\
(Hasinger \etal\  1986). Further study showed that six of the persistently
bright low-mass X-ray binaries (LMXB) form a distinct group: the Z-sources
(Hasinger \&\ van der Klis, 1989. Also see the review by van der Klis, 1995).

As the energy spectrum varies in time, a Z source traces out a characteristic,
Z-shaped track in a X-ray colour-colour diagram (CD) whose  x-axis is the
ratio of the count rates in two low-energy bands while the y-axis is the ratio
of the count rates in two high-energy bands.  The three branches of the Z,
from top to bottom, are called the horizontal branch (HB), normal branch (NB),
and flaring branch (FB).

The fast timing behaviour (millisecond to tens of seconds) is well correlated
to the position of the source along its track, thus defining the state of the
source. Two\footnote{We do not discuss the recently discovered kilohertz QPO,
from Z and Atoll sources or the 45\,Hz QPO (probably HBO) in \sco\,  found
with RXTE (see review by van der Klis 1997)  because they are undetectable with
EXOSAT or Ginga data} distinct types of low  frequency  QPO have been
identified. The horizontal branch oscillations (HBO) with frequencies  between
15 and 55\,Hz have been observed in \cyg, \gx5, \bgx340, \agx17, and \sco. The
$\sim$6\,Hz normal branch oscillations (NBO) are observed from all Z sources.
As \sco\ moves from the NB to the FB the QPO frequency increases rapidly but
smoothly (Priedhorsky \etal\ 1986, Dieters \&\ van der Klis 1999). Apparently
similar changes in frequency are seen for \agx17\ (Penninx \etal\  1990). This
smooth transition in QPO frequency suggests that the NBO and the flaring
branch oscillations (FBO) are related phenomena. On the FB, the QPO frequency
is well correlated with X-ray count rate and position along the Z-track. 
Other Z sources such as \gx5\ and \cyg\ have relatively inconspicuous flaring
branches with, so far, no FBO\footnote{ The QPO on the FB of \cyg\ (Kuulkers
\&\ van der Klis 1995) are thought to be of a different character than the
NBO/FBO of \sco\ and \agx17}. 

Even though the QPO cannot be directly observed with EXOSAT or Ginga, it is
possible to measure the QPO properties as a function of energy using Fourier 
techniques applied to large amounts of high time resolution multichannel data. 
We call the fractional rms variability amplitude as a function of energy the
``rms amplitude spectrum".  By using Fourier cross-spectra we can measure the
phase or, equivalently, the time delay between the QPO at different energies.
We call the  phase difference or time delay between signals as a function of
photon energy the ``phase-delay spectrum" and ``time-delay spectrum",
respectively. 

Cross spectral techniques were first applied to the HBO of \cyg\ by van der
Klis \etal\ (1987). They found that the HBO at higher energies was delayed
(lagged) with respect to the HBO at lower energies by several milliseconds.
More recently, Vaughan \etal\ (1994a) have found that the phase lag increases 
smoothly with increasing energy for the HBO of \gx5.

Ginga data from \cyg\ when it was on its NB (Mitsuda \&\ Dotani 1989) showed
that the NBO in the high energy bands \mbox{(7--18\,keV)} had time-lags
relative to the low-energy bands {\mbox{(1--7\,keV)} of up to $\sim$80\,ms or
equivalently 150\degree\ phase lag. This phase-lag jump occurred near 6\,keV,
where there is also a minimum in the rms amplitude spectrum. 

Vaughan \etal\ (1994a) introduced a new, more sensitive, method for measuring
the phase-delay spectrum. Using this method it was found that there is a sharp
phase jump of $\sim$150\degree\ at $\sim$\,3.5\,keV in the NBO of \gx5\
(Vaughan \etal\ 1999).  However, there was only marginal evidence for a
minimum in the rms amplitude spectrum at this energy. 

The combination of a \about180\degree\ phase shift and a minimum in the rms
amplitude spectrum of the NBO at the same photon energy can be explained by  a
radiation-hydrodynamic model (Lamb 1989, Fortner \etal\ 1989, Miller \&\ Lamb
1992). At near Eddington  accretion rates an approximately spherical flow
forms.  Within this flow a global radiation-hydrodynamic mode can be excited
with a frequency inversely proportional to the inflow time. Because of the
high luminosities, radiation forces become important and the infall time-scale
is lengthened. Consider a density enhancement at the outer boundary; as it
reaches the inner boundary its interaction with the outgoing radiation
increases resulting in an increase or decrease in the  radiation flux reaching
the outer boundary and hence induces, via the change in radiation force, a
density change in the infalling material. Whether the flux increases or
decreases depends upon the relative importance  of the induced luminosity or
Compton scattering opacity changes.  If the influence of the changes in the
Compton scattering optical depth within the radial flow dominates the
influence of the changes in the luminosity during the NBO and the Compton
temperature of the X-ray spectrum is low enough, then the X-ray spectrum
``pivots" during the oscillations (Lamb 1989, Miller \&\ Lamb 1992). As a
result, the oscillations above the pivot energy are \about180 degrees out of
phase with the oscillations below the pivot energy and the oscillation
amplitude is small near the pivot energy. The pivot energy is most sensitive
to the electron temperature in the radial flow. If instead the effect of the
change in the luminosity during the NBO dominates, the X-ray spectrum may
oscillate but not ``pivot" (Psaltis \&\ Lamb 1999). The luminosity at the
NB-FB junction, where the N/FBO frequency changes abruptly (Dieters \&\ van
der Klis 1999), is identified with the Eddington critical luminosity (Hasinger
1987). The change in QPO as \sco\ moves into and along the FB may be explained
either by the   shrinking of the radial flow region (Lamb, 1989) or the
excitation of  non-radial modes (Miller \&\ Park 1995).

We applied the Vaughan \etal\ (1994a) method to EXOSAT data from the NB of
\cyg\ and on the NB and FB of \sco. Ginga data from the FB of \sco\ were also
analyzed. Because instrumental dead-time can mimic a 180\degree\ phase  lag
between low- and high-energy fluxes, it is important to treat it  carefully.
The dead-time was treated differently for the EXOSAT than for the Ginga data.
For Ginga data the dead-time was corrected in the time domain (Vaughan \etal\
1999) whereas for the EXOSAT data the  dead-time effects were corrected in the
frequency domain (van der Klis \etal\ 1987).

Here we report, for the first time, the detection of near 180\degree\ phase
lags in the NBO of \cyg\ with EXOSAT data. This is a direct confirmation of
the results of Mitsuda \&\ Dotani (1989), who used Ginga data.  We set 
approximate limits on the change in pivot energy with position on the NB. No
\sco\ data have previously been analyzed for phase lags. We find for both the
NB and FB of \sco\ that the rms amplitude spectrum increases steeply with
photon energy, with no indication of a minimum. We detect only small phase
lags with no indications of any $\sim$180\degree\ phase lags on either the NB
or FB. 

\section{Observations}

We used data from the ME detectors on-board EXOSAT (Turner \etal\ 1981, White
\&\ Peacock 1988) and from the LAC detectors on the Ginga  satellite (Makino
\etal\ 1987, Turner \etal\ 1989).   The EXOSAT  HER7 observing mode provided 4
energy channels of data with 4 or 8\,ms time resolution. \sco\ was observed
using the HER7 mode on the NB and FB on Aug 25 (day 237) 1985 (Priedhorsky
\etal\ 1986) and on the NB on March 13 (day 072) 1986 (Hasinger \etal\ 1989). 
The energy bands were 0.9--3.1, 3.1--4.9,  4.9--6.6, and 6.6--19.5\,keV. These
\sco\ observations were non-standard in that one half-array of 4 detectors was
off-set pointed ($\sim$12\% collimator response), gathering HER7 data from
only the argon (1--20\,keV) detectors of the ME instrument, while the other
half-array was pointed directly at \sco\ gathering high-time-resolution (HTR3
mode), single-channel (4 or 8\,ms, 5--35\,keV) data from only the xenon
detectors of the ME instrument.

During each NB observation of \sco\ the QPO frequency was generally stable.
Any short frequency excursions ($>$1\,Hz away from the average) were excluded
from the data to keep the average QPO peak narrow. A total of 3040\,s and
5504\,s of data were selected from data taken during the Aug 1985 and March 1986
observations. Both observations spanned the lower normal branch;
1.6$\le$\,\Sz\,$\le$1.9\footnote{The Z-track position is parameterized by \Sz,
the arc length along a smooth Z-track drawn through the points on a
colour-colour diagram, scaled such that \Sz=1 at the HB/NB vertex and \Sz=2 at
the NB/FB vertex (see Dieters \&\ van der Klis 1999).} . The QPO frequency was
6.41$\pm$0.05\,Hz (full width half maximum, FWHM, of 1.91$\pm$0.14\,Hz) during
the Aug 1985 observation and 6.15$\pm$0.05\,Hz (FWHM of 2.4$\pm$0.4\,Hz)
during the Mar 1986 observation. The EXOSAT Aug 25, 1985 observation also
contained data while \sco\ was on the flaring branch. These data were grouped
according to QPO frequency. Only in the frequency range 14.5--15.5\,Hz were
there sufficient data for phase-lag studies; i.e., a total of 5632\,s spanning
2.06$\le$\,\Sz\,$\le$2.17.

The Ginga data on \sco\ studied here were obtained using MPC-3 mode data,
which consists of 12 spectral channels over the energy range  1.5--18.7\,keV
(high detector gain) with 8\,ms time resolution. During these observations on
March 9/11 1989, \sco\ was on the FB (Hertz \etal\ 1992). Data (3320\,s) were
selected from the lowest segment of the flaring branch where the QPO frequency
was 14.5\,Hz\,$\pm0.3$\,Hz (FWHM of 6.3$\pm$1.3\,Hz). This QPO frequency is
found at Z track positions corresponding to 2.06$\le$\,\Sz\,$\le$2.09. There
are no high-time- resolution Ginga data available when \sco\  showed NBO.

The EXOSAT observations of \cyg\ were obtained on November 14/15  (day
318/319) of 1985 using the HER7 mode with a time resolution of 4\,ms.  During
this observation which lasted $\sim$13\,hr, \cyg\ was  on the NB and FB
(Hasinger \&\ van der Klis 1989,  Kuulkers \etal\ 1996). Four energy bands
(from the argon layer detectors) were available:  0.9--3.1, 3.1--4.7,
4.7--6.3, and 6.3--19.7\,keV (i.e., roughly the same energy bands as used for
\sco).  The NBO are clearly present over the range \Sz=\,1.3--1.7.   We
divided the NB into two parts: in the ranges 1.1$\le$\,\Sz\,$\le$1.6  (``upper
NB'') and 1.6$\le$\,\Sz\,$\le$2.0  (``lower NB''). This yielded 14048\,s and
15168\,s of observations for the upper NB and the lower NB, respectively. The
mean QPO frequency was 5.6$\pm$0.1\,Hz with FWHM 2.2$\pm$0.3\,Hz on the upper
NB and  5.8$\pm$0.1\,Hz with FWHM 1.7$\pm$0.5\,Hz on the lower NB.

\section{Analysis}

We first divided the data into 32\,s long data segments and then calculated
the individual complex Fourier transforms (CFTs). This was done for each
energy channel. Then,  the cross spectrum was  calculated as the product of
the CFT of one channel with the complex conjugate of the CFT of the other
channel  for each possible pair of energy channels.  In each data set the
cross spectra of all data segments were averaged together.  The average cross
spectrum thus consists of a sequence of individual cross vectors, one at each
Fourier frequency. Further averaging was done in frequency to produce an
average cross-vector over the QPO frequency range for each pair of energy
channels.

Instrumental dead time induces an anti-correlation between energy channels
that appears in the cross correlation function as an anomalously large
negative value at zero lag, and in the cross spectrum as a purely negative
real part in each cross vector that is independent of Fourier frequency,
corresponding to a phase difference of 180\degree. Both Poisson fluctuations
and luminosity signals induce such phase lags (Lewin \etal\ 1988).  This
effect is known as channel cross-talk. The channel cross-talk induced by the
source signal (QPO) was found to be insignificant compared to that from the
Poisson fluctuations (Vaughan \etal\ 1999) for typical QPO amplitudes and
count-rates as measured with EXOSAT and Ginga. 

As discussed by Vaughan \etal\ (1999), there are three possible approaches to
handling the dead-time effects induced by Poisson fluctuations. First, a
``signal-free'' cross vector obtained from the data itself in a frequency range
devoid of signals can be subtracted from the signal (QPO) cross vector. Second,
by knowing the dead-time process, the modified probability distribution in each 
channel can be calculated and hence from it the induced covariance and induced
dead-time cross vector, which can then be subtracted from the QPO cross vector.
Last, the observed count rates  (as affected by dead-time) can be converted to
counts per live-time and then the corrected rates can be used in calculating the
complex Fourier spectra and so the cross spectra are without the
dead-time-induced negative, real component.  These three methods are referred to
as the frequency domain, statistical and time domain methods respectively. We
used the time domain (last) method for the Ginga data and the frequency domain
(first) method for the EXOSAT data.

The dead-time process for EXOSAT is complicated (Andrews 1984, Andrews \&\
Stella 1985, Tennant 1987, Berger \&\ van der Klis 1994) and not entirely
understood. Therefore it is not possible to estimate the effects of dead-time
on the count-rate distribution with the required degree of precision. A
time-domain count rate correction is not possible because at high count rates
the dependence of dead-time on count-rate is not known precisely enough, and
also because the count rates, especially in the high-energy channel are too
low ($\le$1 per 4\,ms time bin). There are too many zero count bins for an
accurate conversion. We therefore adopted the procedure of subtracting the
average real part of the observed cross spectrum at high frequencies 
(72--128\,Hz) from the cross vector at each frequency of the cross  spectrum.
This procedure is  similar to that used by van der Klis \etal\ (1987). The
underlying assumption is that there is no intrinsic signal in the
high-frequency region of the cross spectrum. At these frequencies,
high-frequency noise is the only known source signal. Its amplitude, when QPO
are detectable, is small  at high frequencies since its cut-off frequency is
between 60 -- 80\,Hz in the NB and near 35\,Hz on the FB (Dieters \&\ van der
Klis 1999). Thus at high frequencies any time lags should be due to
instrumental effects and so should be a good measure of the dead-time induced
channel cross-talk.  The imaginary part of the high-frequency vector should be
ignored because the dead-time-induced vector is purely real.  In all cases we
found a large negative real vector, and the distribution of the imaginary
part, at high frequencies, was consistent with a null-mean normal distribution
(a large number of estimates were averaged, so the central limit theorem
applies) with a variance due only to counting statistics.

For Ginga data the dead-time process is better understood and, moreover, count
rates in all energy channels are high enough to correct them in the time
domain. We used the known, fixed dead-time per event to correct the counts in
each time bin and in each energy channel to counts per live time, before
computing cross spectra (Mitsuda \&\ Dotani 1989, Vaughan \etal\ 1999).

After calculating the cross spectra and applying the above dead-time 
correction strategies, the average signal cross vectors were calculated for
each energy channel pair by averaging over the frequency range of the QPO
peak. The signal frequency range used was 
$\nu$$_{c}$$\pm$$(3/4)$$\Delta\nu$\,Hz, where $\nu_c$ is the QPO  centroid
frequency and $\Delta\nu$ is its FWHM.  For the EXOSAT data,  the error
estimates in the average signal cross vectors are based upon the observed
variance in the individual values of the real and imaginary components of both
the high-frequency vector and the QPO vector. These error estimates are
consistent with the error expected from counting statistics, given the  count
rates, rms amplitude of the QPO, and length of the  observations and assuming
perfect coherence between channels for the QPO. For the Ginga data the errors
were directly estimated from the cross vectors given the number of frequencies
and Fourier transforms used (equation 11 of Vaughan \etal\ 1994a). 

Because QPO are weak compared with counting (Poisson) noise, measuring phase
differences involves first setting a detection threshold on the magnitude of
the average ``signal'' cross vector and then, only in the case of significant
detection, setting confidence limits on the phase.  Conceptually and 
computationally, the procedure is similar to detecting and measuring  coherent
pulsations (Vaughan \etal\ 1994b). 

Each of the terms averaged together in estimating the average cross vector of
the QPO is the product of two complex random variables whose real and
imaginary parts each have a Gaussian distribution with a null mean.  In the
absence of a source signal, the real and imaginary parts of the average cross
vector are themselves distributed as null-mean Gaussian random variables and
the squared magnitude of the average cross spectrum, $|G|^2$, is distributed
as $\exp(-|G|^2/|G_0|^2)$, where  $|G_0|^2$ can be well estimated from the
variances in the real and imaginary parts of the cross spectrum. The phase
angle (argument) of G is distributed uniformly over $[-\pi,+\pi]$.

A significant detection is when the magnitude of the signal cross vector is
greater than some previously chosen detection threshold. We chose a 95\%
confidence detection threshold for all cross spectra. If the average cross
vector fails to exceed the detection threshold, we can conclude nothing about
the presence or absence of phase differences (angle of cross vector) or place
any meaningful limits on their size. Either the QPO signal is too weak
compared with the noise background to allow a significant detection, or the
signals in the different channels were not sufficiently coherent (Vaughan
\etal\ 1994a). 

Once a significant detection of the cross vector is made, the phase delay (the
argument or phase angle of the signal cross vector) can be  found. We
determined the 90\% confidence intervals of the phase angle. Naturally the
phase delay can be zero when the cross vector is  significant. In this case we
can set a meaningful upper limit on the  phase delay. 

\begin{figure}
{\psfig{file=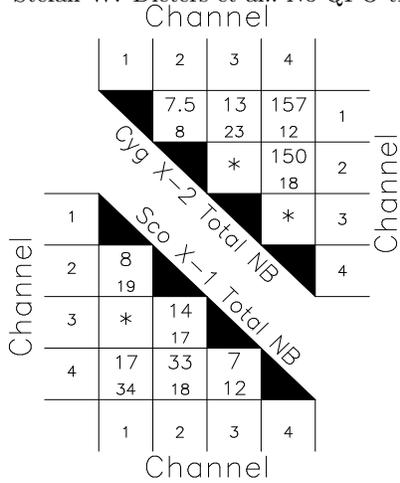,height=5.5cm,width=8.8cm,angle=-90}}
{\caption[]{The channel by channel results for the 
cross-spectral analysis of the EXOSAT data for \cyg\ and \sco. The upper
panel is for the NBO of \cyg\ from data covering both the upper and
lower sections of the NB. The lower panel is for the NBO of \sco\
using the sum of the NB data from 25 Aug 1985 and 13 March 1986. 
In each box the upper number is the measured phase lag in degrees
while the lower number is the 90\% uncertainty. A {\large $\ast$}
indicates that the magnitude of the average cross vector is 
{\it not} significant.}}
\end{figure}

For both the EXOSAT and Ginga data we used the $\chi$$^2$ method of Vaughan
\etal\ (1994a), which makes use of all available information by taking into
account all statistically significant cross-vectors between each possible
combination of channels. Similar but less  significant results were obtained
by only using the cross vectors between the lowest energy channel and each of
the other channels.

Towards the lowest and highest energies measured by EXOSAT and Ginga the PHA
spectrum of both sources drops steeply. At low energies this is because of 
interstellar absorption while at high energies it is due to a combination of
the steep intrinsic energy spectrum and a drop in detector efficiency. Thus the
central energies, particularly of the lowest and highest, and necessarily
broadest channels, used to measure the rms amplitude and phase do not
reflect the energy of most of the photons being measured. We can define the
photon-average central energy as the energy which splits the number of 
photons evenly within the channel's range. The photon average central  energy
of the lowest and highest HER7 channels where measured with multi-channel (32
or 128 channel) ME Ar data taken on on Aug 3 (day 216) 1984 and Jun 26/27 (day
177/178) 1985, respectively. All 12 channels of the Ginga MPC-3 data were used
to find these limits for \sco\ on the FB. These half-count-rate energies are
2.3 and 9.6\,keV, 2.9 and 12.0\,keV for the EXOSAT and Ginga data for \sco\,
respectively, and 2.4 and 7.0\,keV for the EXOSAT \cyg\ data.
 
Although, to first order, the theoretical rms amplitude spectrum (Miller and
Lamb 1992) can be directly compared with our observed rms spectrum, to do a
comparison good enough to place limits on pivot energy  requires folding a
fine grid of models through the detector matrix. Modeling the observed phase
lag spectrum would require, in addition to knowing the PHA spectrum,
assumptions about the variation of rms amplitude and phase with energy within
each energy channel. Given the current state of our knowledge we only provide
general statements about the energy of any phase jump.

\section{Results}

\subsection{\cyg.}

For \cyg\ (EXOSAT data) we determined time/phase delays in three  independent
frequency intervals, one centered on and covering the NBO 
($\nu_{NBO}$$\pm$$\frac{3}{4}$$\Delta\nu$, i.e.\, 4.2--7.2\,Hz) and two
(1.2--4.2 and 7.2--10.2\,Hz) on each side of the NBO peak in the power
spectrum. In the frequency ranges adjacent to the NBO we found no significant
cross vectors between any pair of energy channels. 

In the NBO frequency range, particularly on the upper normal branch, we found
several significant cross vectors which showed a large phase/time delay
between the two highest channels (3 and 4) and no significant delays between
the lower energy channels (Fig.\,2).  The actual channel by channel
detections, for the whole NB, are given in the upper part of Fig\,1.

\begin{figure}
\psfig{file=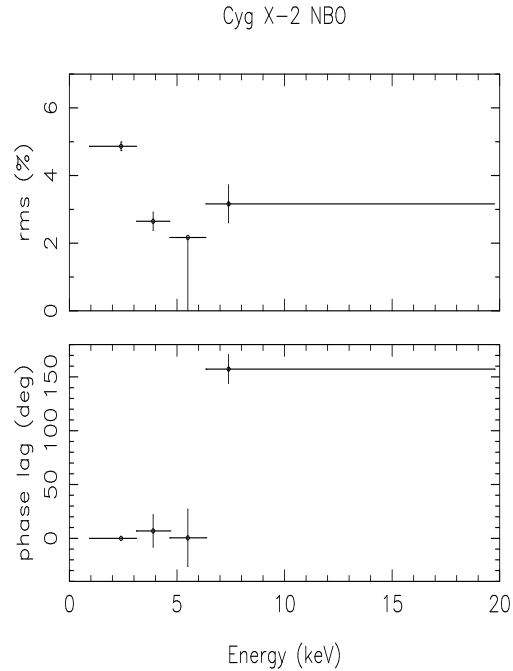,height=9.0cm,width=8.8cm,angle=-90}
\caption[]{The rms amplitude spectrum and phase delay spectrum from
EXOSAT data for the NBO of \cyg, averaged over the part of the NB where 
the NBO is detectable. A 2$\sigma$ upper limit is shown for the rms 
amplitude in channel 3. The 1$\sigma$ limit is 0.5\%\,rms.
The phase delays are with respect to channel 1, so this channel
has no error bar.}
\end{figure}

We determined a mean rms amplitude (energy) spectrum of the NBO.  We fixed the
NBO centroid frequency at 5.7\,Hz and the FWHM at 2.0\,Hz, the values obtained
from a fit to the power spectra of the sum of channels 1--4, and allowed only
the strength of the NBO to vary in our fits to the power spectra of the
individual channels. There is  a dip at channel 3 (4.7--6.3\,keV) in the rms
amplitude spectrum  (2$\sigma$ upper limit in Fig.\,2). 

We find that the rms amplitude spectra and the phase delay spectra are the 
same within the $1\sigma$ errors on the upper and lower branch.  Since the
dips in the model rms amplitude spectra are comparable in width
($\sim$1.8\,keV) to the width of channel 3, a change in pivot energy to an
adjacent energy bin; i.e. to the middle of channel 2 (3.9\,keV) or to the
photon average central energy of channel 4 (7\,keV) would cause a large change
in the rms amplitude spectrum. So we conclude that the pivot energy did not
change more than $\sim$$\pm1.5$\,keV along the NB. As a comparison with the
models  (Miller \& Lamb 1992); a change in electron temperature from 0.5 to
1.0\,keV changes  the pivot energy from 5.2 to 7.3\,keV and so can be
excluded.
 
In the upper NB the higher energy time/phase delay amounted to 78$\pm$4\,ms or 
equivalently 161$\pm$9\degree. In the lower NB the time/phase lags were  very
similar, i.e.; 70$\pm$6\,ms or alternatively 144$\pm$12\degree. Averaged over
the  whole NB the time/phase delays were 75$\pm$4\,ms or 154$\pm$7\degree.
Considering the source energy spectrum and model rms amplitude spectra, the
pivot energy should reasonably lie between (approximately) the middle of
channel 3 (5.5\,keV) and photon average central energy of channel 4 (7\,keV).
We therefore conclude that the pivot energy is near 6\,keV. 
 
Our results confirm, using another instrument (the EXOSAT ME), the existence
of both a near-180\degree\ phase jump and a rms amplitude minimum in \cyg\ as
previously seen by Mitsuda \&\ Dotani (1989) using Ginga LAC data. Because the
instrumental dead-time effect produces a 180\degree\ phase delay, it is
reassuring that the  results found with two different instruments are in
complete agreement. 

\subsection{\sco.}

\begin{figure}
{\psfig{file=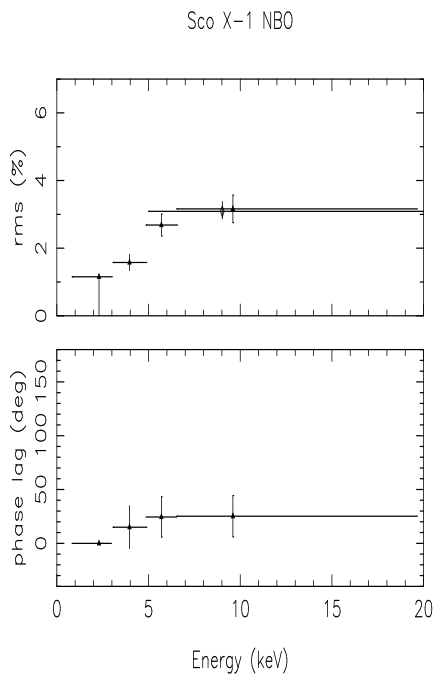,height=9.0cm,width=7.5cm,angle=-90}
\caption[]{The rms amplitude spectrum and phase delay spectrum from
EXOSAT data on the NBO of \sco, averaged over the Aug 25, 1985 and March 13,
1986 data sets. The phase delays are with respect to channel 1 on so 
this channel has no error bar.}}
\end{figure}

For the NB EXOSAT data from \sco, we determined the phase/time lags  over the
frequency range spanning the NBO peak
($\nu_{NBO}$$\pm$$\frac{3}{4}$$\Delta\nu$)  and also over several other
frequency ranges for each data set (Aug 25 1985 and March 13 1986) separately.
We found no significant cross-vectors for any of the non-QPO frequency ranges.
The data, particularly from 13 March 1986 (day 072), were of sufficient
quality to detect significant QPO cross vectors. Our channel-by-channel
detections for the sum of the NB data are listed in the lower part of Fig.\,1.
At the 90\% confidence level we found no phase delays in the NB greater than
60\degree\ (largest limit in both data sets) between any pair of energy
channels. By combining both data sets, an overall upper limit of 42\degree\
(highest possible phase lag using the $\chi$$^{2}$ method) at the 90\% 
confidence level can be set on any phase jump on the NB between  
$\sim$2.3~and~9.6\,keV (photon average central energies of channels 1 and 4
respectively).  The phase delay spectrum is shown in the lower panel of
Fig.\,3. The separate rms amplitude spectra and also the average of both
observations (upper panel of Fig.\,3) show no sign of a dip. 

\begin{figure}
{\psfig{file=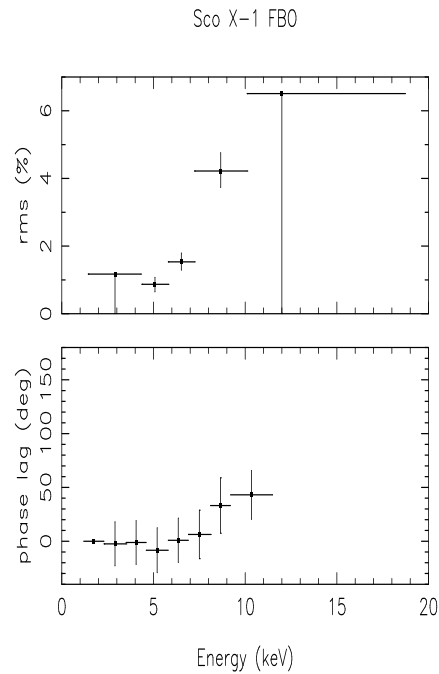,height=9.0cm,width=7.5cm,angle=-90}
\caption[]{The rms amplitude spectrum and phase delay spectrum from
Ginga data for the FBO of \sco. The phase delays are with respect to 
channel 1 on so this channel has no error bar. When a phase delay is 
not significant it has been omitted from the plot. Some correlation in 
phase lag between adjacent bins is introduced by the $\chi$$^{2}$ method.}}
\end{figure}

The EXOSAT FB data were of sufficient quality only to measure the phase delay
between channels 2 and 3. In the 10--20\,Hz range covering the QPO, this delay
was consistent with zero (90\% confidence upper limit of 22\degree). No
significant cross-vectors were found for other frequency ranges. These tests
included frequencies typical of NBO.

The Ginga FB data showed significant cross vectors between all channels,
except those at the lowest and highest energies (lower panel of Fig.\,4) where
the QPO signal is very weak, as can be seen in the rms energy spectrum of the
upper panel of Fig.\,4. Again, the phase delays were consistent with zero. The
90\%\ confidence upper limit on any phase lag between channels 1 to 8  (energy
range 2.9--12\,keV) is  88\degree\ (between 4.5 and 8.1\,keV this value is
51.4\degree).  The rms energy spectrum on the FB has a similar form to that on
the NB, with no evidence for a dip. 

\section{Discussion}

Our methods are capable of detecting a phase jump in EXOSAT data  as is
evidenced by our clear detection of such a jump for \cyg, confirming  the
result of Mitsuda \& Dotani (1989).  For \sco\ we find no evidence for a phase
jump or minimum in the rms amplitude spectrum in the range 2--10\,keV on the
NB, or 3--12\,keV on the FB, with 90\%\ confidence upper limits on the size of
such a phase jump of 42 and 88 degrees, respectively. Either \sco\ does not
have a rocking NBO spectrum or  the pivot energy is either
$\stackrel{<}{_{\sim}}$\,2\,keV or $\stackrel{>}{_{\sim}}$\,10\,keV. In view
of the fact that the rms amplitude is steeply increasing with energy much like
the expected rms spectrum above the pivot energy (Miller \&\ Lamb 1992), we
favour the lower possible pivot energy. This is clearly different from the
pivot energy of \cyg\ which is close to 6\,keV (Mitsuda \&\ Dotani 1989, this
paper) and \gx5\  at about 3.5\,keV (Vaughan \etal\ 1999). 

In many respects all the Z sources are very similar. In particular the
behaviour of the NBO is nearly identical between sources. An exception may be
\cgx349\ (Kuulkers \&\ van der Klis, 1998).  So it is initially a surprise that
the pivot energy should be so different (factor 2-3). This is particularly the
case for \cyg\ and \gx5 which based upon source behaviour are considered more
alike than either compared to \sco. 

Nearly all behavioural properties of the Z sources are solely a function of Z
track position. Several arguments identify the mass accretion rate as
governing the Z track position. Also  the nature of the accretion flow is
apparently dependent on \mdot\ relative to the Eddington accretion rate
\mdot$_{Edd}$. A comparison between Z sources (Dieters \&\ van der Klis 1999)
has shown that the NBO become detectable at  much the same Z track position
suggesting that \Sz\ is an accurate measure of  \mdot relative to
\mdot$_{Edd}$ across Z sources.

We considered as did Vaughan \etal\ (1999) whether the different pivot
energies in different sources are due to sampling a pivot-energy/\mdot\ 
relation at different mass accretion rates. This seems unlikely because  the
estimated \mdot's are comparable  for all Z sources in the same spectral
branch (Hasinger \&\ van der Klis 1989, Hasinger \etal\ 1990, Vrtilek \etal\
1990). More sensitively, \mdot/\mdot$_{Edd}$ must be very similar to produce
NBO (Lamb 1989, Fortner \etal\ 1989, Miller \&\ Lamb 1992). Confirming that
changes in \mdot\ on the NB make little difference is our observation that the
pivot energy does not change greatly ($\pm$1.5\,keV) along the NB of
\cyg. Our NB observations of \sco\ were made at very similar Z track positions
as the lower NB observation for \cyg. Also, for \sco\ the phase-lag and
rms-energy spectra are similar on the NB and FB despite the radically
different flow conditions associated with sub (NB) and super (FB) critical
Eddington accretion. 

Based upon differences in the HB and FB morphologies and the relative
strengths of the FBO there seem to be  two groups, of Z sources: those like
\cyg\ and those like \sco. These two sources may represent the extremes in a
continuum of properties. The Sco-like sources are \sco, \agx17, and \cgx349\
and the Cyg-like sources are \cyg, \gx5, and \bgx340. Both inclination angle
(Hasinger \&\ van der Klis 1989, Kuulkers \etal\ 1994,1996) and neutron star
magnetic field strength (Psaltis \etal\ 1995, Psaltis \&\ Lamb 1999) have been
put forward as the underlying factor giving rise to these two groups.

The fact that phase jumps are observed in \gx5\ and \cyg\ and not in \sco\ is
consistent with this division into two groups. If this subdivision can indeed
be applied to phase delays in Z sources,  then in the energy range above
$\sim$2\,keV we expect to see a phase jump in the NBO of \bgx340, whereas no
such phase jump is expected for \agx17\ and \cgx349.

The rms amplitude spectrum of \agx17\ shows no dip between $\sim$5\,keV and
$\sim$12\,keV on either the normal or flaring branches (Penninx \etal\  1990).
Similarly there is no evidence for a dip in the NBO rms amplitude spectrum of
\cgx349\ above $\sim$5\,keV, and there is no phase lag/lead between low 
(1--5\,keV) and high (5--10\,keV) photons $\ge$80\degree/118\degree\ (90\%
confidence) (Ponman \etal\ 1988). These measurements are consistent with both
sources  being Sco-like. In the case of \bgx340, a Cyg-like source the rms
amplitude spectrum shows no dip at $\stackrel{>}{_{\sim}}$5\,keV. A dip such
as  that of \cyg\ is  probably excluded but this source may be like \gx5 with
a low pivot energy.  The above evidence is broadly consistent with the pivot
energy also  being related to the division of Z-sources into two groups.

If, pivot energy depends upon inclination angle, say through asymmetries in the
QPO production region caused by the presence of an inner ``puffed-up'' torus,
then this could explain the differences between \cyg\ and \sco. However, the
pivot energies of \cyg\ and \gx5\ are reversed with the assumed ranking of
sources  by inclination angle within the \cyg\ group as proposed by  Kuulkers
\&\ van der Klis (1995).

In the radiation-hydrodynamic model (Lamb 1989, Fortner \etal\ 1989, Miller
\&\ Lamb 1992), the N/FBO  are produced in a relatively cool, spherically
symmetric, radial inflow region that surrounds a hotter inner corona and
magnetosphere of the neutron star. The assumption of approximately spherical
symmetry eliminates inclination as causing the differentiation of the
Z-sources. Recent extensions to the basic model have suggested that the
neutron star  magnetic field strength is the controlling factor (Psaltis
\etal\ 1995, Psaltis \&\ Lamb 1999).
 
If the NB/FBO is caused by a radiation force - opacity feedback mechanism,
where the opacity is primarily due to Compton scattering, then the small
changes in optical depth during each oscillation will cause the energy
spectrum to pivot (Lamb 1989). As a natural consequence, there is a minimum in
the rms amplitude spectrum of the QPO and about a 180\degree\ phase jump  at
the pivot energy in the phase lag spectrum. This appears to be the case for
\cyg.  The longer escape time of photons scattered downward in energy by the
cool radial flow may explain why the observed phase shift is not exactly 180
degrees (Lamb 1989). 
 
The pivot energy depends mainly, though weakly, upon the electron temperature
(T$_{\rm e}$) of the cool radial flow and to a lesser  extent on its optical
depth $\tau$. The electron temperature  can be estimated fairly accurately by
computing the Compton temperature of the X-ray spectrum we observe. The
possible values for the optical depth are limited by the twin constraints of
the observed spectrum and the lack of any X-ray pulsations. In the
 simplified spectral model of Miller \&\ Lamb (1992) 
the value of T$_{\rm e}$ needed to fit the data of Mitsuda \&\ Dotani (1989)
and presented here for the \cyg\ NBO is the same as the value of T$_{\rm e}$
computed from the observed spectrum of \cyg. Given the range of Compton
temperatures determined from the X-ray spectra of Z-sources (0.5--1.5\,keV),
the range of possible  pivot energies is limited to those shown in Figure 5 of
Miller \&\ Lamb (1992), i.e. 5--10\,keV. This is incompatible with the
observed energy of the phase jump ($\sim$3.5\,keV) of \gx5\ and the implied
pivot energy ($\le$2.3 or $\ge$9.6\,keV) for \sco. 

If, instead of optical depth modulations dominating during NBO, the effect of
the luminosity changes dominates, then the X-ray spectrum may oscillate
without pivoting (Psaltis \&\ Lamb 1999). Modeling of the shapes of the Z
tracks of various sources (Psaltis \etal\ 1995) requires that the neutron
stars in the Sco-like subclass of Z sources  have weaker magnetic fields than
the neutron stars in the Cyg-like subclass. A weaker magnetic field will
produce a softer  radiation spectrum  from the magnetosphere and hot central
corona. The shape and luminosity of this softer spectrum is affected more by
changes in the mass flux than the spectrum emerging from a stronger field
neutron star, so luminosity changes may dominate NBO production, accounting
for the lack of any evidence for a pivoting spectrum in \sco.

However, this does not explain why \gx5\ has  a phase-jump (pivot) energy
lower than the range compatible with the models of Miller \&\ Lamb (1992) used
successfully for \cyg.  

\subsection{Conclusion}

There are two possible sets of explanations for the results presented here for
\sco\ and \cyg\ and for those on \gx5\ (Vaughan \etal\ 1999). First; there 
may be asymmetries in the NBO producing region causing the pivot  energy to
depend upon viewing angle (i.e., inclination).

Second, within the context of the radiation-hydrodynamic models for the
NBO/FBO, the QPO  of \cyg\ and \gx5\ are dominated by optical depth changes
while those of \sco\ are dominated by luminosity variation, the underlying
cause of the differences in QPO production being due to the  neutron star of
\sco\ having weaker  magnetic field. Both explainations are at this stage 
unable to account for all details of the distribution of the measured pivot
energies over the sources.
 
\begin{acknowledgements}We wish to thank Guy Miller who supplied, in tabular
form, the results of NBO models as presented in Miller and Lamb (1992) and  
Demetrios Psaltis for valuable comments. This work was supported in part by
the Netherlands Organization for Scientific  Research (NWO) under grant PGS
78-277. At the Max Planck Institute for Astrophysics S.\,Dieters was supported
by grant ERB-CHRX-CT93-0329 of the European Commission (HCM program). Brian
Vaughan acknowledges support from NASA  under grant NAG\,5--3293. 
\end{acknowledgements} 

\rf{Andrews D., 1984, EXOSAT Express 5, 31.}
\rf{Andrews D., \&\ Stella L., 1985, EXOSAT Express 10, 35.}
\rf{Berger M., \&\ van der Klis M., 1994, A\&A 293, 175.}
\rf{Dieters S.W., \&\ van der Klis M., 1999, MNRAS, in press, astro-ph/9909472}
\rf{Fortner B., Lamb F.K., \&\ Miller G.S., 1989, Nat 342, 775.}
\rf{Hasinger G., 1987, A\&A 186, 153.}
\rf{Hasinger G., \&\ van der Klis M., 1989, A\&A 225, 79.}
\rf{Hasinger G., Langmeier A., Sztjanjo M., \etal\ 1986, Nat 319, 469.}
\rf{Hasinger G., Priedhorsky W.C., \&\ Middleditch J., 1989, ApJ 337, 843.}
\rf{Hasinger G., van der Klis M., Ebisawa K., Dotani T., \&\ Mitsuda~K.,
 1990, A\&A 235, 131.}
\rf{Kuulkers E., \&\ van der Klis M., 1995, A\&A 303, 801.}
\rf{Kuulkers E., \&\ van der Klis M., 1996, A\&A 314, 567.}
\rf{Kuulkers E., \&\ van der Klis M., 1998, A\&A 332, 845.}
\rf{Kuulkers E., van der Klis M., Oosterbroek T., \etal\ 1994, A\&A 289, 795.}
\rf{Kuulkers E., van der Klis M., Vaughan B.A., 1996, A\&A 311, 197.}
\rf{Lamb F.K., 1989 in: Two Topics in X-ray Astronomy, 23rd
ESLAB Symp., J. Hunt \&\ B. Battrick (eds.) 1989, ESA SP-296, v1, p. 215.}
\rf{Lewin W.H.G., van Paradijs J., van der Klis M., 1988,
Sp. Sci. Rev. 46, 273.}
\rf{Makino F. and ASTRO-C Team, 1987, Astr. Lett. Comm. 25, 233.}
\rf{Middleditch J., \&\ Priedhorsky W.C., 1986, ApJ 306, 230.}
\rf{Miller G.S., \&\ Lamb F.K., 1992, ApJ 388, 541.}
\rf{Miller G.S., \&\ Park, M-G., 1995, ApJ 440, 771.}
\rf{Mitsuda K., \&\ Dotani T., 1989, PASJ  41, 557.}
\rf{Penninx W., Lewin W.H.G., Mitsuda K., \etal\ 1990, MNRAS 243, 114.}
\rf{Priedhorsky W., Hasinger G., Lewin W.H.G., \etal\ 1986, ApJ 306, L91.}
\rf{Ponman T.J., Cooke B.A., \&\ Stella L., 1988, MNRAS 231, 999.}
\rf{Psaltis D., Lamb F.K. \&\ Miller G.S. 1995, ApJ 454, L137.}
\rf{Psaltis D., \&\ Lamb F.K. 1999, in preparation.}
\rf{Tennant A.F., 1987, MNRAS 226, 963.}
\rf{Turner M.J.L., Smith A., \&\ Zimmermann H.U., 1981, Sp. Sci. Rev. 30, 513.}
\rf{Turner M.J.L., Thomas H.D., Patchett B.E. \etal\ 1989, PASJ 41, 345.}
\rf{van der Klis M., Jansen F., van Paradijs J. \etal\ 1985, Nat 316 225.}
\rf{van der Klis M., 1995, In {\it X-ray Binaries}, W.H.G. Lewin, J. van Paradijs, \& E.P.J. van den Heuvel (eds.), Cambridge Univ. Press, Great Britain, pp252-307.}
\rf{van der Klis M., Hasinger G., Stella L., \etal\ 1987, ApJ 319, L13.}
\rf{van der Klis M., 1997, in NATO/ASI Ser, The Many Faces of Neutron Stars, ed
R. Buccheri, J. van Paradijs, \& M.A. Alpar (Dordrecht: Kluwer), astro-ph/9710016}
\rf{Vaughan B.A., van der Klis M., Lewin W.H.G., \etal\ 1994a, ApJ 421, 738.}
\rf{Vaughan B.A., van der Klis M., Wood K.S., \etal\ 1994b, ApJ  435, 362.}
\rf{Vaughan B.A., van der Klis M., Lewin W.H.G., \etal\ 1999, A\&A 343, 197.} 
\rf{Vrtilek S.D., Raymond J.C., Garcia M.R., \etal\ 1990, A\&A 235, 162.}
\rf{White N.E., \&\ Peacock A., 1988, in: X-ray Astronomy with
EXOSAT, R. Pallavicini \&\ N.E. White (eds.), Mem. S. A. It.
59, 7.}

\end{document}